%% file: author.tex
\documentclass[graybox]{svmult}

\usepackage{mathptmx}       % selects Times Roman as basic font
\usepackage{helvet}         % selects Helvetica as sans-serif font
\usepackage{courier}        % selects Courier as typewriter font
\usepackage{type1cm}        % activate if the above 3 fonts are
\usepackage{makeidx}         % allows index generation
\usepackage{graphicx}        % standard LaTeX graphics tool
\usepackage{multicol}        % used for the two-column index
\usepackage[bottom]{footmisc}% places footnotes at page bottom
\usepackage{url}
\usepackage[numbers]{natbib}
\usepackage{aas_macros}
\usepackage{enumerate}

\newcommand{\photz}{photo-\emph{z}}
\newcommand{\photzs}{photo-\emph{z}s}

\makeindex             % used for the subject index
\begin{document}

\title*{Using gamma regression for photometric redshifts of survey galaxies}
\author{J.~Elliott, R.~S.~de~Souza, A.~Krone-Martins, E.~Cameron, E.~E.~O.~Ishida, J.~Hilbe}
\institute{J.~Elliott \at Harvard-Smithsonian Center for Astrophysics, 60 Garden Street, Cambridge, MA 02138, USA, Max-Planck-Institut f{\"u}r extraterrestrische Physik, Giessenbachstra\ss e 1, 85748, Garching, Germany, \email{jonathan.elliott@cfa.harvard.edu}
\and R.~S.~de~Souza \at MTA E\"otv\"os University, EIRSA ``Lendulet'' Astrophysics Research Group, Budapest 1117, Hungary
\email{rafael.2706@gmail.com}
\and A.~Krone-Martins \at SIM, Faculdade de Ci\^encias, Universidade de Lisboa, Ed. C8, Campo Grande, 1749-016, Lisboa, Portugal
\email algol@sim.ul.pt
\and E.~Cameron \at Department of Zoology, University of Oxford, Tinbergen Building, South Parks Road, Oxford, OX1 3PS, United Kingdom
\email dr.ewan.cameron@gmail.com
\and E.~E.~O.~Ishida \at Max-Planck-Institut f\"ur Astrophysik, Karl-Schwarzschild-Str. 1, 85748 Garching, Germany
\email emille@mpa-garching.mpg.de
\and J.~Hilbe \at Arizona State University, 873701,Tempe, AZ 85287-3701, USA, Jet Propulsion Laboratory, 4800 Oak Grove Dr., Pasadena, CA 91109, USA
\email j.m.hilbe@gmail.com
\and for the COIN collaboration
}

\maketitle

\vspace{-3.5cm}
\abstract{Machine learning techniques offer a plethora of opportunities in tackling \emph{big data} within the astronomical community. We present the set of Generalized Linear Models as a fast alternative for determining photometric redshifts of galaxies, a set of tools not commonly applied within astronomy, despite being widely used in other professions. With this technique, we achieve catastrophic outlier rates of the order of $\sim1\%$, that can be achieved in a matter of seconds on large datasets of size $\sim1,000,000$. To make these techniques easily accessible to the astronomical community, we developed a set of libraries and tools that are publicly available.}

\vspace{-0.5cm}
\section{Introduction}
\label{sec:1}
\vspace{-0.5cm}

Generalized Linear Models~\citep[GLMs;][]{nel72} are widely used throughout other scientific disciplines such as: biology \citep{bro93}, medicine \citep{lindsey1999}, and economics \citep{pin98}, and is available within the overwhelming majority of contemporary statistical software packages. However, they have been very little used within the astronomical community

There are plenty of opportunities to apply GLMs within astronomy, and one particularly important problem is the estimation of photometric redshifts (\photz{}). 

Galaxy spectra are made up of many of its physical properties, including  morphology, age, metallicity, star formation history, merging history, and a host of other confounding factors in addition to its redshift. This makes robust estimation of \photz{}s a difficult task. Estimation is usually done in two ways, by template fitting, or by using machine learning techniques.

There exist several studies that have investigated the advantages of the publicly available codes that estimate the \photz{} of galaxies~\citep[for a glimpse on the diversity of existent methods, see][and references therein]{hildebrandt2010}. The overall performance of most codes is good, demonstrating catastrophic errors from 5 to 9\%, which is considered reliable within the field. There are also a number of growing techniques that implement a hybrid approach of template and machine learning techniques~\citep{Carrasco14a}.

Despite the current advancement within this field, there still exist large practical difficulties. In the next years there are a large number of surveys that will start having \emph{big data} catalogues, e.g., the \textit{Large Synoptic Survey Telescope}\footnote{\url{http://www.lsst.org/lsst}} \citep{lsst},  \textit{EUCLID}\footnote{\url{http://sci.esa.int/euclid}} \citep{euclid} or the \textit{Wide-Field Survey Infrared Telescope}\footnote{\url{http://wfirst.gsfc.nasa.gov}} \citep{wfirst}. Current techniques will become difficult to employ if they require large training sets, and as such, this warrants the need for fast and reliable \photz{} methods that are capable of robustly estimating redshifts quickly, and on large training datasets.

We introduce GLMs as a new technique to quickly and robustly estimate galaxy \photz{}s. We show that it can run in a matter of seconds on a single core computer, even for millions of objects. As part of the COsmostatistics INitiative (COIN\footnote{\url{https://asaip.psu.edu/organizations/iaa/iaa-working-group-of-cosmostatistics}}) collaboration, we created and distributed easy to use software, and web-applications for use of the wider community in estimating \photz{}s\footnote{\url{https://github.com/COINtoolbox}}.

\vspace{-0.5cm}
\section{Methodology}
\label{sec:2}
\vspace{-0.5cm}

We will not go into the details of deriving GLMs or the formula that is used in our technique, we instead encourage the reader to see the details in~\citet{Elliott15a} and references therein. However, we note the importance of GLMs, such that they allow you to choose the type of distribution you want to model. GLMs are applicable to the entire set of  \textit{exponential families} of distributions: Gaussian/normal, gamma, inverse Gaussian, Bernoulli, binomial, Poisson, and negative binomial. For example, in this study, we want to predict the \photz{}s of galaxies from multi-wavelength photometry. For such a study, the gamma distribution is favourable, as a redshift is positive and continuous, as is the gamma distribution. To then use the gamma distribution to predict redshifts, we utilised the following machine learning methodology:

\begin{enumerate}[1.]

  \item The data was randomly split into training and test sets.
  
  \item Robust principal component analysis (e.g., \citealt{Candes11a,desouza2014}) was carried out on the complete data set.
 
  \item We utilised a gamma family distribution to reflect the fact that measured redshifts are positive and continuous. 
  
  \item The predicted \photz{} for the test data was calculated using the principal component projections of the test data set and the best-fit GLM using the training sample.
  
  \item To measure how well the \photzs{} were estimated, we employed a metric commonly used in the literature, specifically, the catastrophic error.
  
\end{enumerate}

\vspace{-0.5cm}
\section{Data Samples}
\label{sec:3}
\vspace{-0.5cm}

We used two publicly available galaxy datasets to test the technique outlined in the previous section. The first was the \textit{PHoto-z Accuracy Testing} (PHAT), an international intitiative to identify the most promising \photz{} methods. We used their publicly available simulated datasets that contains $169,520$ simulated galaxies with redshifts ranging from $z=0.02-2.24$, and magnitudes in 11 filters ($u$, $g$, $r$, $i$, $z$, $Y$, $J$, $H$, $K$, $IRAC1$, and $IRAC2$).

Given that this dataset was purely synthetic, we also used a real dataset acquired from the {\it Sloan Digital Sky Survey} \citep[SDSS;][]{York00a}. For details on the query see~\citet{Elliott15a}. The sample used contained $1,347,640$ galaxies with a redshift range of $z=0-1.0$, with magnitudes in 5 filters ($u'$, $g'$, $r'$, $i'$, and $z'$). Comparisons with dereddened values showed no inconsistencies

\vspace{-0.5cm}
\section{Results}
\label{sec:4}
\vspace{-0.5cm}

\begin{figure}[h]
	\includegraphics[width=9cm]{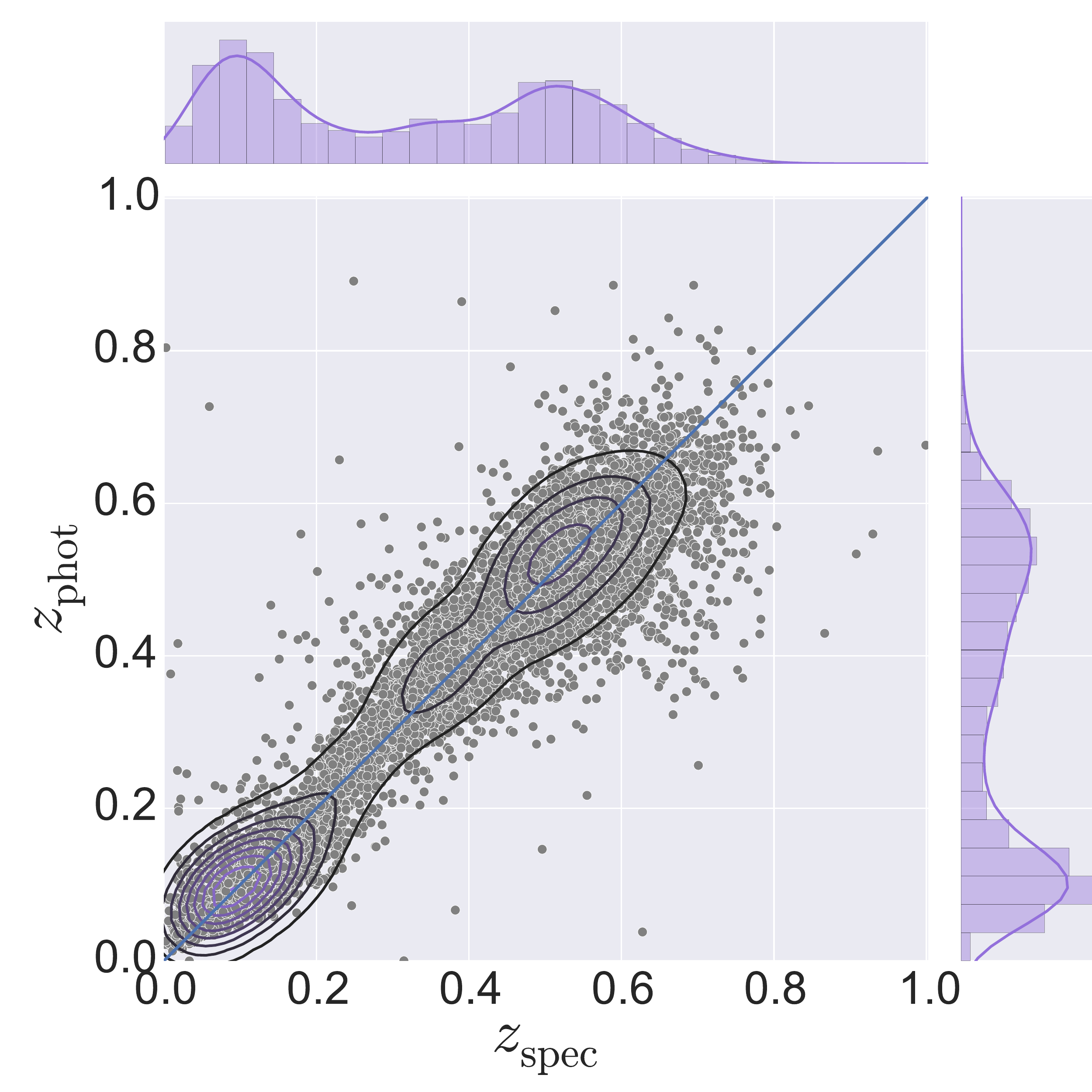}
   \caption{The 2D probability density of the predicted redshift from the GLM fit vs. the spectroscopic redshift (central plots). The upper and right subplots in each panel depict the redshift distribution along \photz{} and $z_{\rm spec}$, respectively.
   }
  \label{fig:kde_plot_2d}
\end{figure}

Both data sets were fit using an AMD Athlon X2 Dual-Core QL-64 processor with $1.7\, \rm GB$ RAM on the Ubuntu  10.04 operating system, which represents an old laptop at today's standards. We achieved catastrophic errors of $1.4\%$ for the PHAT0 data set and $8\%$ for the SDSS data set, within $\sim1200$, and $10$ seconds. Lower catastrophic errors of $\sim1\%$ could be achieved when using more principal components, but would take longer computational time, $\sim5000\, \rm s$. We plot the best-fit GLM models for SDSS datasets in Fig.~\ref{fig:kde_plot_2d}.

\vspace{-0.5cm}
\section{Conclusions}
\label{sec:5}
\vspace{-0.5cm}

The astronomical community has left Generalized Linear Models relatively untouched, despite its use throughout the academic world. We have demonstrated their ease of use and quick applicability to estimate the \photz{}s of galaxies. This technique has been shown to be competitive with current techniques implemented, that can require larger training sets and longer time for their algorithms to learn. Such properties of this technique will become important in the close future when upcoming wide field sky surveys, such as the LSST, will start collecting data at enormous rates per day.

To make GLMs more accessible to the community, we developed a set of software libraries written in Python and R, that can be easily implemented into people's own work. A web application is also available to be used instantly without the need for installation.

\begin{acknowledgement}
We thank V. Busti, E. D. Feigelson, M. Killedar, J. Buchner,  and A. Trindade for interesting suggestions and comments. 
JE, RSS and EEOI thank the SIM Laboratory of the \emph{Universidade de Lisboa} for hospitality during the development of this work. 
Cosmostatistics Initiative (COIN)\footnote{\url{https://asaip.psu.edu/organizations/iaa/iaa-working-group-of-cosmostatistics}} is a non-profit organisation whose aim is to nourish the synergy between astrophysics, cosmology, statistics and machine learning communities.
This work was partially supported by the ESA VA4D project (AO 1-6740/11/F/MOS). AKM thanks the Portuguese agency \emph{Funda\c c\~ao para Ci\^encia e Tecnologia} -- \emph{FCT}, for financial support (SFRH/BPD/74697/2010). EEOI is partially supported by the Brazilian agency CAPES (grant number 9229-13-2). Work on this paper has substantially benefited from using the collaborative website AWOB developed and maintained by the Max-Planck Institute for Astrophysics and the Max-Planck Digital Library. This work was written on the collaborative \texttt{WriteLatex} platform, and made use of the GitHub a web-based hosting service and \texttt{git} version control software. This work made use of the cloud based hosting platform \texttt{ShinyApps.io}. This work used the following public scientific Python packages \texttt{scikit-learn} \texttt{v0.15}~\citep{Pedregosa11a}, \texttt{seaborn v0.3.1}, and \texttt{statsmodels v0.6.0}. Funding for SDSS-III has been provided by the Alfred P. Sloan Foundation, the Participating Institutions, the National Science Foundation, and the U.S. Department of Energy Office of Science.
\end{acknowledgement}

\input{referenc}
\end{document}

%% file: referenc.tex
%%%%%%%%%%%%%%%%%%%%%%%% referenc.tex %%%%%%%%%%%%%%%%%%%%%%%%%%%%%%
% sample references
% %
% Use this file as a template for your own input.
%
%%%%%%%%%%%%%%%%%%%%%%%% Springer-Verlag %%%%%%%%%%%%%%%%%%%%%%%%%%
%
% BibTeX users please use
\bibliographystyle{plainnat}
\bibliography{ref}